\documentclass[10pt, conference]{IEEEtran}

\usepackage{hyperref}
\usepackage{cite}
\usepackage{graphicx}
\usepackage{float}
\usepackage{pgfplots}
\usepackage{caption}
\usepackage{subcaption}
\usepackage{listings}
\usepackage{svg}
\usepackage{amsmath}
\usepackage{pdfpages}
\usepackage{pgfplotstable}
\usepackage{multirow}

\usepackage{pdfcomment}

\begin{document}
\bstctlcite{IEEEexample:BSTcontrol} % To force the use of "et al" in references
\graphicspath {./images/} 

\title{\texttt{AVEC}: Accelerator Virtualization in Cloud-Edge Computing for Deep Learning Libraries}

\author{\IEEEauthorblockN{Jason Kennedy, Blesson Varghese and Carlos Rea\~{n}o}
\IEEEauthorblockA{Queen's University Belfast, UK\\
\{jkennedy49, b.varghese, c.reano\}@qub.ac.uk}
}

\maketitle

\thispagestyle{plain}
\pagestyle{plain}

\begin{abstract}
Edge computing offers the distinct advantage of harnessing compute capabilities on resources located at the edge of the network to run workloads of relatively weak user devices. This is achieved by offloading computationally intensive workloads, such as deep learning from user devices to the edge. Using the edge reduces the overall communication latency of applications as workloads can be processed closer to where data is generated on user devices rather than sending them to geographically distant clouds. 
Specialised hardware accelerators, such as Graphics Processing Units (GPUs) available in the cloud-edge network can enhance the performance of computationally intensive workloads that are offloaded from devices on to the edge. The underlying approach required to facilitate this is virtualization of GPUs. 
This paper therefore sets out to investigate the potential of GPU accelerator virtualization to improve the performance of deep learning workloads in a cloud-edge environment. 
The \texttt{AVEC} accelerator virtualization framework is proposed that incurs minimum overheads and requires no source-code modification of the workload. 
\texttt{AVEC} intercepts local calls to a GPU on a device and forwards them to an edge resource seamlessly. 
The feasibility of \texttt{AVEC} is demonstrated on a real-world application, namely OpenPose using the Caffe deep learning library.
It is observed that on a lab-based experimental test-bed \texttt{AVEC} delivers up to 7.48x speedup despite communication overheads incurred due to data transfers.
\end{abstract}


\IEEEpeerreviewmaketitle

\section{Introduction}
\label{sec:introduction}
\begin{figure}[t]
  \centering
  \includegraphics[width=0.47\textwidth]{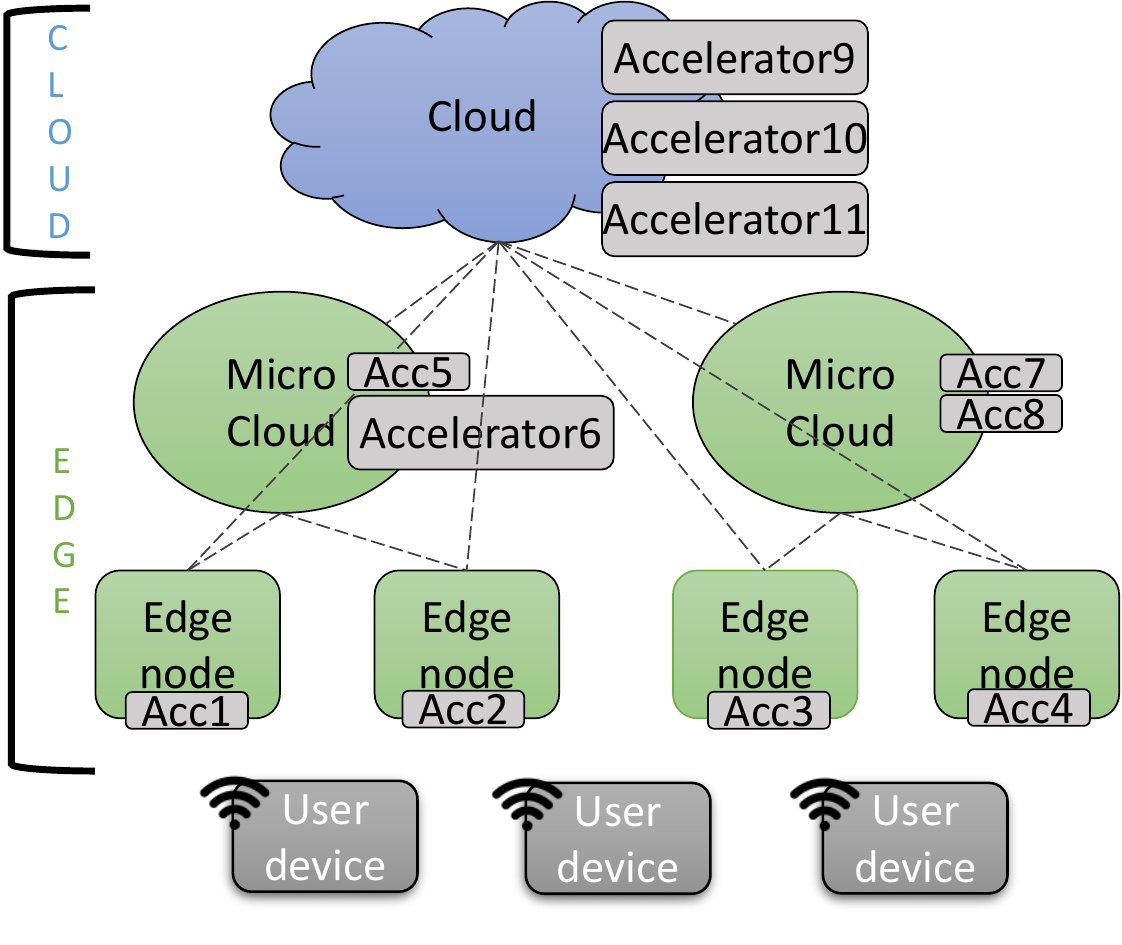}
  \caption{An example cloud-edge architecture}
  \vspace{-0.5cm}
  \label{fig:CloudEdge}
\end{figure}

Edge computing offers multiple layers of resources at the edge of the network between an end-user device and the cloud~\cite{arch-01, edgecomputing-03}. The premise of edge computing is to bring workloads closer to where data is generated for minimising communication latencies between devices and geographically distant clouds and for reducing the ingress bandwidth demand to the cloud~\cite{enorm, dyverse}. Edge resources may be employed to offload workloads from the cloud to the edge or from devices to the edge. Edge resources may vary in form factor depending on where they are located. For example, a home router may be augmented with computing resources to be an edge node or alternatively a dedicated micro-cloud with more computing resources may be employed. Figure~\ref{fig:CloudEdge} provides an exemplar of an edge computing architecture~\cite{Fog}. 

Workloads that execute on the edge, such as deep learning~\cite{DeepLearning, scission} can benefit from hardware accelerators that provide massive parallelism on relatively small form factor processors. One example is the graphics processing unit (GPU) that this paper focuses on. Alternate accelerators include FPGAs (field-programmable gate arrays) or TPUs (tensor processing units). Co-locating accelerators on the edge may reduce the execution time of workloads thereby making it a more compelling proposition for the edge. Workloads will need to typically rely on frameworks, such as CUDA~\cite{CUDA} or OpenCL\cite{OpenCL} to exploit the parallelism offered by GPUs. The GPU executes functions that are referred to as kernels, which are routines stored on the accelerator and are separate from those executed on the CPU. The CPU will send an input to the accelerator for the kernel and then receive the kernel output.

All end user devices may not have immediate access to an edge node with a GPU since all edge resources will not be GPU powered. Therefore it is anticipated that workloads will need access to GPUs that are located on other resources in the cloud-edge network. For example, an accelerator that is hosted in a micro-cloud rather than on an edge node. This is facilitated by the virtualization technology; accessing remote resources and using them as if they were local resources. The cloud-edge architecture mentioned above envisions that resources in a cloud-edge network will be able to access virtualized accelerators hosted on nodes in the network. Thus, in practice, resources with no physical accelerators can potentially access accelerators on other nodes. 

\begin{figure}[t]
  \centering
  \includegraphics[width=0.4\textwidth]{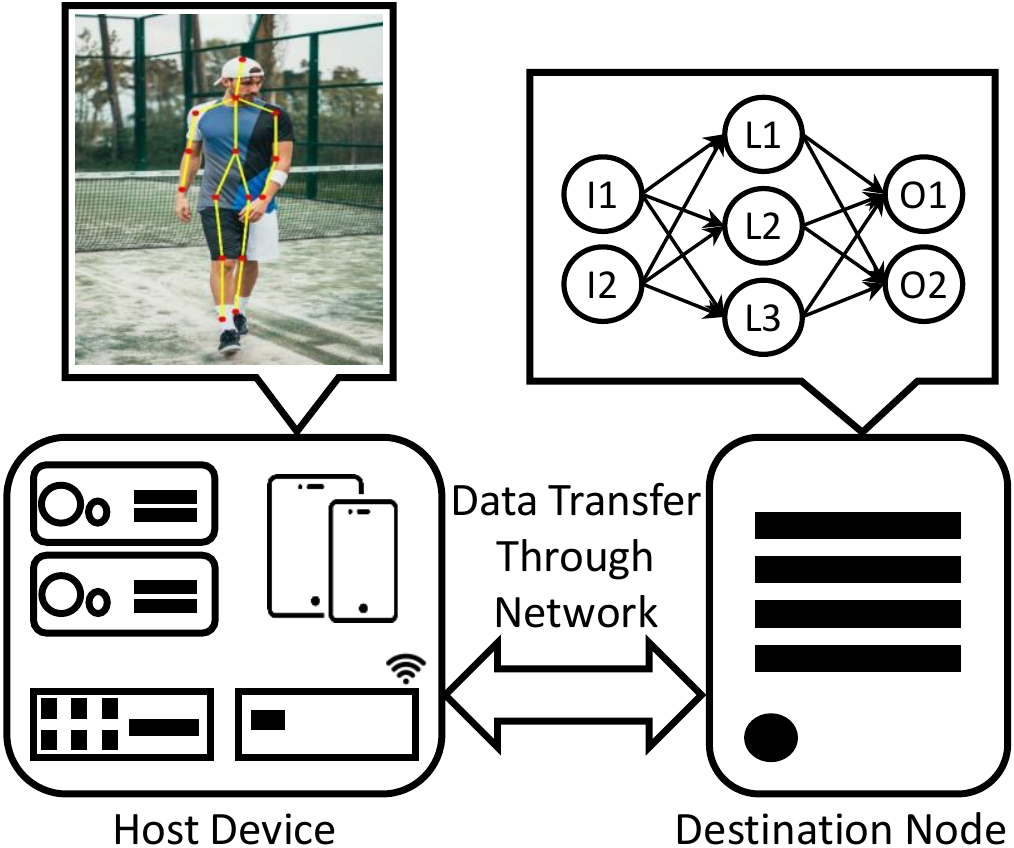}
  \caption{Example of accelerator virtualization in cloud-edge}
  \vspace{-0.5cm}
  \label{fig:Example}
\end{figure}

GPU virtualization in the cloud-edge network can be achieved by intercepting local calls to a GPU and forwarding them to a resource hosting an accelerator, executing the GPU kernel on the virtualized accelerator, and returning the output of execution. This offers the possibility for a computationally weaker device to send its data through the network so that a resource hosting a GPU can execute the workload instead of the device (a high-level view is provided in Figure~\ref{fig:Example}).

Although there is significant research on virtualizing GPUs for high-performance computing systems, such as rCUDA~\cite{rCUDAVirtual} or gVirtuS~\cite{Gvirtus}, and on GPUs for low-power embedded computers suited for the edge~\cite{ReanoICPP18, MontellaCCPE17}, there is currently no research that focuses on virtualization in a cloud-edge network. Therefore, the research reported in this paper investigates the \textit{potential of GPU accelerator virtualization to improve the performance of deep learning workloads in a cloud-edge environment}. More specifically, the following research questions are addressed:

\textbf{Q1}: \textit{How can accelerator virtualization be offered at the edge given its limited compute resources?} To address this question, this paper will present a novel prototype \texttt{AVEC} (accelerator virtualization in cloud-edge computing), a framework that supports the use of virtual GPU accelerators in the cloud-edge continuum. \texttt{AVEC} is portable and can be employed in heterogeneous cloud-edge environments. The key advantage of the proposed framework is that it does not require source code modification. 

\textbf{Q2}: \textit{What overheads are incurred in virtualizing accelerators at the edge?} The key motivation in addressing this question is to determine whether the computational benefits of having access to more powerful GPUs are offset by additional overheads incurred due to communication. It is observed that the proposed \texttt{AVEC} framework can deliver up to 7.48x speedup despite communications overheads incurred due to data transfer to a remote GPU.
    
\textbf{Q3}: \textit{Can GPU accelerator virtualization at the edge improve the execution performance of real-world applications?} In this paper, we select a deep learning library, namely Caffe~\cite{caffe} to demonstrate this. We show that it is feasible to offload computationally intensive components of workloads that require GPUs from weak devices and edge nodes to nodes hosting GPUs via virtualization. The experimental evaluation is carried out in the context of a deep learning application which uses the aforementioned Caffe library, namely OpenPose~\cite{OpenPose}.

The main research contributions of this paper are as follows: 
\begin{itemize}
    \item The development of \texttt{AVEC}, a first prototype of a low overhead and edge performance enhancing framework that supports the use of virtual GPU accelerators in the cloud-edge continuum.
    \item The demonstration of the benefits of accelerator virtualization at the edge by transparent execution of the popular Caffe deep learning library kernels using a real-world use case on remote GPUs. 
\end{itemize}
This research confirms the feasibility of offloading workloads to virtualized GPU accelerators in cloud-edge environments.

The remainder of this paper is organised as follows. Section~\ref{sec:LiteratureReview} presents existing techniques implementing virtualization and management of resources. Section~\ref{sec:Motivation} highlights the motivation and the necessity for implementing accelerator virtualization within a cloud-edge network. Section~\ref{sec:SolutionProposed} proposes \texttt{AVEC}; the accelerator virtualization framework and provides its implementation. Section~\ref{sec:Experiments} presents the results obtained from experimental studies. Finally, Section~\ref{sec:Conclusion} concludes this paper by considering future work.

\section{Related Work}
\label{sec:LiteratureReview}
Existing literature presents accelerator virtualization solutions in the context of large-scale systems, such as high-performance computing (HPC) clusters and clouds, and low power embedded devices.

\subsection{Accelerator Virtualization in Large-scale Systems}

There are multiple solutions available for general-purpose computing on graphics processing units (GPGPU) virtualization in HPC clusters. These solutions are based on middleware libraries or remote procedure calls (RPC). 

Remote CUDA (rCUDA)~\cite{rCUDAVirtual} is one middleware-based solution that virtualizes remote GPUs. The middleware intercepts CUDA calls on a client node and forwards them to a server node hosting a physical GPU. This allows for the GPU accelerator to be logically decoupled from the physical node, thereby allowing for other clients to access and share the same physical GPU on the server. A similar framework that is developed for OpenCL is VOCL~\cite{VOCLVirtual}. 

vCUDA~\cite{vCuda} is another CUDA based accelerator virtualization solution for HPC clusters. This solution is RPC-based rather than middleware to achieve accelerator virtualization. GViM~\cite{gvim} is an early implementation of GPU virtualization in which multiple virtual machines are hosted on a single physical node and access the same physical GPU. GViM achieves virtualization by using API interception, so that the CUDA function calls from an application running on a virtual machine, can be intercepted and sent to the host machine in a privileged domain for execution.

Alternate solutions for GPU virtualization include DS-CUDA~\cite{DS-CUDA} and Grid-CUDA~\cite{liang2011gridcuda}.
DS-CUDA is a GPU virtualization solution designed for use in the cloud. This solution incorporates a redundancy mechanism by mapping two physical cloud accelerators on to a single virtual accelerator. The outputs of both accelerators are compared; if the two results do not match, DS-CUDA automatically re-runs the CUDA API calls until the same result is seen. This mechanism improves accuracy, but the overheads remain unknown. 

Grid-CUDA makes use of RPC to redirect workloads within a grid of nodes to enable parallel execution. It is noted that the use of RPC incurs large overhead costs, and the use of GRID-CUDA could lower the overall performance.

GVirtuS~\cite{Gvirtus} is another accelerator virtualization solution that is based on virtual machines operating over TCP/IP model, thereby offering remote accelerator virtualization. GVirtuS is independent of the hypervisor. However, the choice of hypervisor affects the performance. The hypervisor allows multiple virtual machines or remote devices to interact with the same physical hardware accelerator. The GVirtuS framework is designed so that the CUDA library interacts with a front-end GPU virtualizer. The back end of GVirtuS deals with the hardware, by unpacking the CUDA library call and then assigning memory for it to be executed. The interception is made using a CUDA wrapper library. 

qCUDA~\cite{qCUDA} is proposed to improve the performance in areas such as bandwidth, for local GPU virtualization. They achieve virtualization by incorporating API interception and is designed for QEMU-KVM hypervisor. Their testing focused on bandwidth performance and they report above 95\% bandwidth efficiency when compared to native execution. The main contribution qCUDA provides is efficient memory transfers. This is made possible by eliminating the extra memory copies between guest and host, by shifting a contiguous guest physical address in memory as a host virtual address in the QEMU process via mapping.

\begin{figure}[t]
  \centering
  \includegraphics[width=0.46\textwidth]{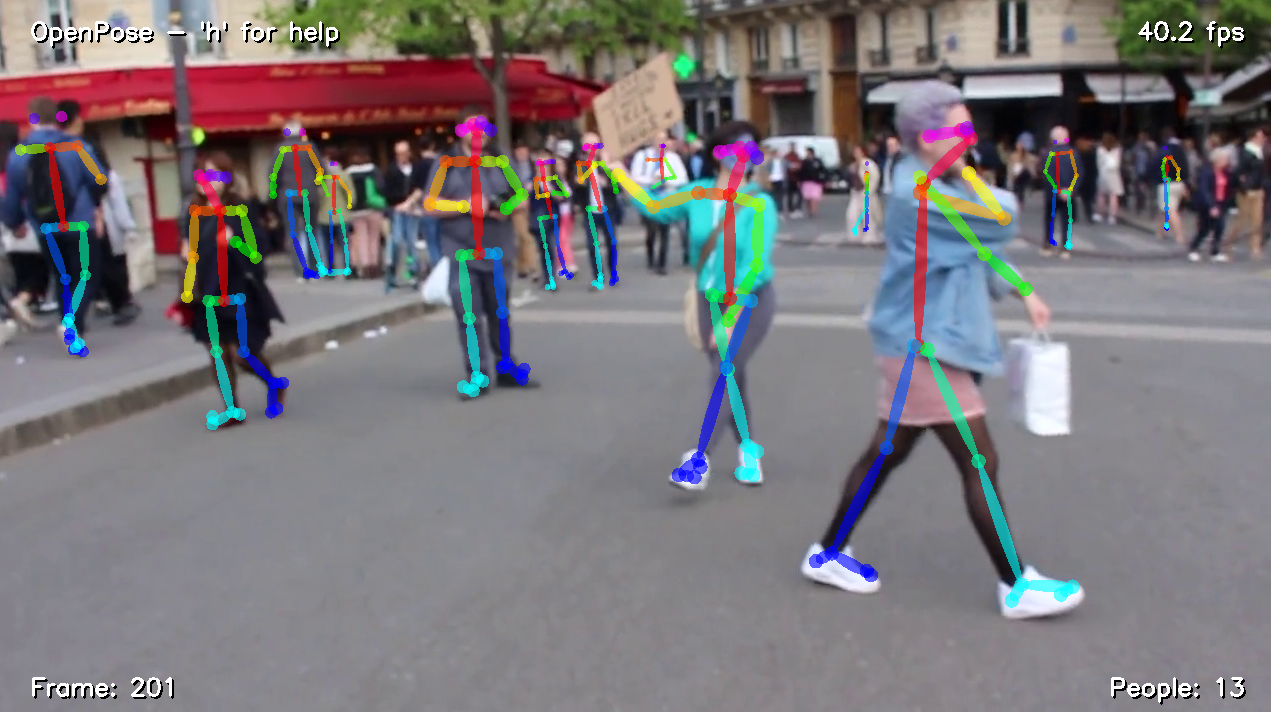}
  \caption{Example output of the OpenPose application}
  \vspace{-0.5cm}
  \label{fig:opDemo}
\end{figure}

\subsection{Accelerator Virtualization on Embedded Devices}

Recently, GVirtuS incorporated accelerator virtualization for ARM based single board computers (SBC)~\cite{GVirtuS-ARM}. This is achieved by intercepting the front end API stub that connects low powered SBCs to GPU accelerators located on x86 servers. The framework offloads workloads from the ARM based SBCs to remote accelerators. Results highlight that the framework is best suited for longer running workloads. This is because the latency factor due to communicating between nodes decreases over time (performance increases as the communication time compared to kernel execution time decreases).

qCUDA-ARM~\cite{qCUDA-ARM} is an extension of the qCUDA framework and incorporates SBCs. This is a first attempt to virtualize GPUs to work on the ARM architecture. While it uses the same conceptual architecture of qCUDA, the memory management process is modified. A key CUDA function used by qCUDA is cudaHostRegister(), which is not supported in ARM devices. As such, the team has implemented memory copying in a different way, so that an additional copy of the pinned memory section is created, mapping this new region to the memory in the guest CUDA application. Although this utilizes twice the amount of memory, it is noted that it accelerates the execution of other memory related CUDA calls such as cudaMemcpy(). The authors report in their experimental findings that they can achieve up to 90\% of the native CUDA speed for pinned memory. Furthermore, they achieve almost native performance on computation bound applications such as matrix multiplication.

RAPID~\cite{RAPID} project is a European Commission funded project that investigates virtualization of lower power SBCs for offloading workloads in a heterogeneous environment. A peer-to-peer sharing mechanism in which devices from smartphones to cloud data centers are connected is envisioned to take advantage of large accelerators in the network to augment the performance of SBCs. 

Alternate low-powered embedded FPGAs are virtualized in existing research~\cite{HyperVisorFPGA, VirtualFPGA}. These FPGAs are designed for specific tasks and can be dynamically and partially reconfigured for other tasks.

In summary, we can conclude that existing accelerator virtualization will not work in a cloud-edge environment for a number of reasons. Firstly, each solution is intended to operate within a specific environment comprising only specific devices. Cloud-edge computing is a heterogeneous environment and this needs to be accounted for. Furthermore, additional overheads can arise with cloud-edge computing, such as latency from wireless communication, that these solutions have not considered. Lastly, the research discussed is designed for virtual machines unlike light-weight deployments, such as containers, employed in a cloud-edge deployment.

\section{Motivation}
\label{sec:Motivation}
% Points to be covered:
% Computationally intensive workloads require hardware acceleration
% Accelerators need to be located at the edge
% Multiple workloads need to share edge accelerators
% Minimum source code modification should be required
% The overheads should not offset the performance benefit

\begin{figure}[t]
  \centering
  \includegraphics[width=0.40\textwidth]{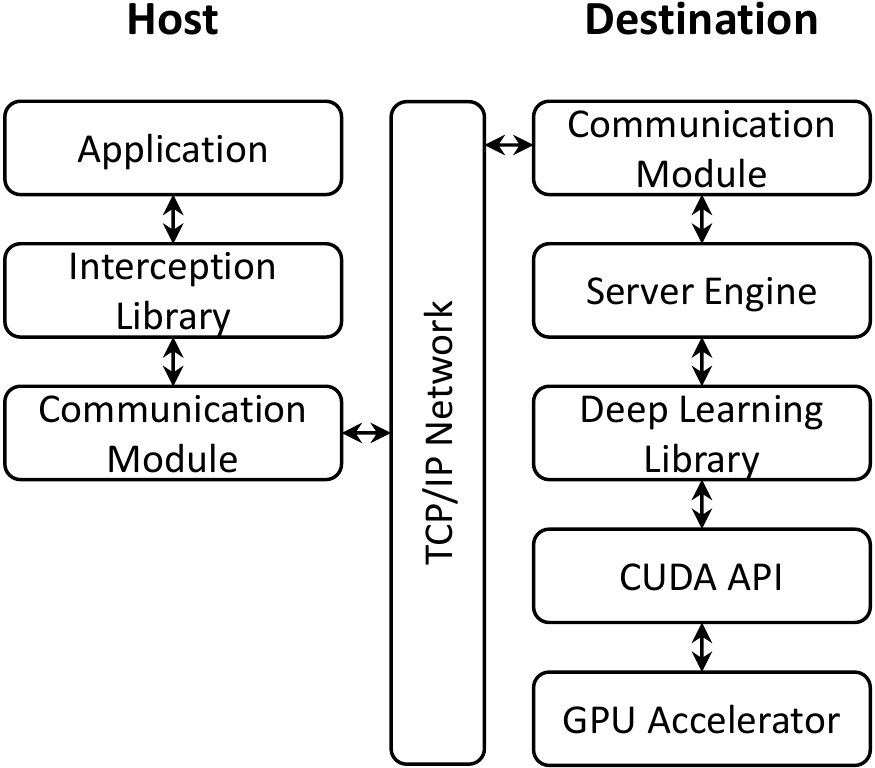}
  \caption{Architecture of the \texttt{AVEC} framework}
  \vspace{-0.5cm}
  \label{fig:architecture}
\end{figure}

The following aspects in relation to design, environment, and usability have been factored in for the accelerator virtualization framework \texttt{AVEC} that is proposed in this paper for a cloud-edge computing environment:

\textit{(1) Computationally intensive workloads require hardware acceleration}~-- Many workloads that will be executed in the cloud-edge environment will be computationally intensive. For example, deep learning is an important class of workloads that has found multiple applications in modern mobile apps~\cite{enorm, dyverse} and smart cities~\cite{Metropolis}. In the cloud-edge environment, these workloads can be partitioned and executed across cloud and edge resources for performance gain~\cite{scission, neurosurgeon}. However, the use of hardware accelerators can enhance the performance of these workloads. In the design of \texttt{AVEC}, the execution of computationally intensive workloads, such as deep learning, have been considered. 

\textit{(2) Accelerators need to be brought to the edge}~-- Although computationally intensive workloads can rely on accelerators in the cloud, there is performance gain when they are located at the edge of the network~\cite{Fog}. A collection of end-user devices can access these accelerators on the edge and enhance the performance of workloads. Or a cloud application can offload selected services to an edge resource hosting accelerators for servicing user requests closer to the source. In both cases, response time of an application can be minimised and more computationally intensive workloads can get executed closer to their source for privacy reasons. In \texttt{AVEC}, processing data nearer to the source has been taken into account. 

\textit{(3) Multiple workloads need to share edge accelerators}~-- The edge environment is anticipated to be busy with users since billions of end-user devices will be connected. Cisco estimated that there would be around 50 billion connected devices by 2020~\cite{Evans2011}. Therefore, it is essential that multiple devices can share an edge accelerator. While moving accelerators to the edge make them more accessible to end devices, solutions designed to make them accessible should be able to execute multiple workloads concurrently. These will need to isolate the user space and provide safe memory access on the hardware accelerator. In this paper, the virtualization solution has been designed to offer isolation of user space in future versions, thus enabling the execution of multiple workloads at the edge.

\textit{(4) Minimum source code modification should be required}~-- An ideal acceleration solution offered at the edge should be less intrusive (i.e. should not require low-level source code modification). Giunta et al.~\cite{GVirtuS-ARM} note the importance of transparency in this manner in their work. Instead the solution should be flexible so that existing distributed applications that leverage accelerators can be scheduled on accelerators whether they are located on an edge node, the micro-cloud or the cloud, based on availability and proximity to the end-user for achieving an agreeable quality of service. This design criteria has been taken into account while designing the \texttt{AVEC} accelerator virtualization framework in which no source code modification is required. The framework intercepts calls to the accelerator from a running application and redirects them suitably to a remote accelerator.

\textit{(5) Overheads should not offset performance benefit}~-- In providing an accelerator virtualization solution overheads will be inevitably incurred~\cite{Fog}. These overheads are due to the additional operations, such as transferring data from an end-user device to GPU kernels that execute remotely and receiving the results of execution back on the device. However, these overheads can be minimised by using efficient data transfer and execution strategies and offsetting the overheads by performance gains. In this paper, the virtualization solution attempts to minimise the overheads and offset them by achieving a performance gain on deep learning workloads.

\section{\texttt{AVEC} Framework}
\label{sec:SolutionProposed}
\begin{figure}[t]
  \centering
  \includegraphics[width=0.42\textwidth]{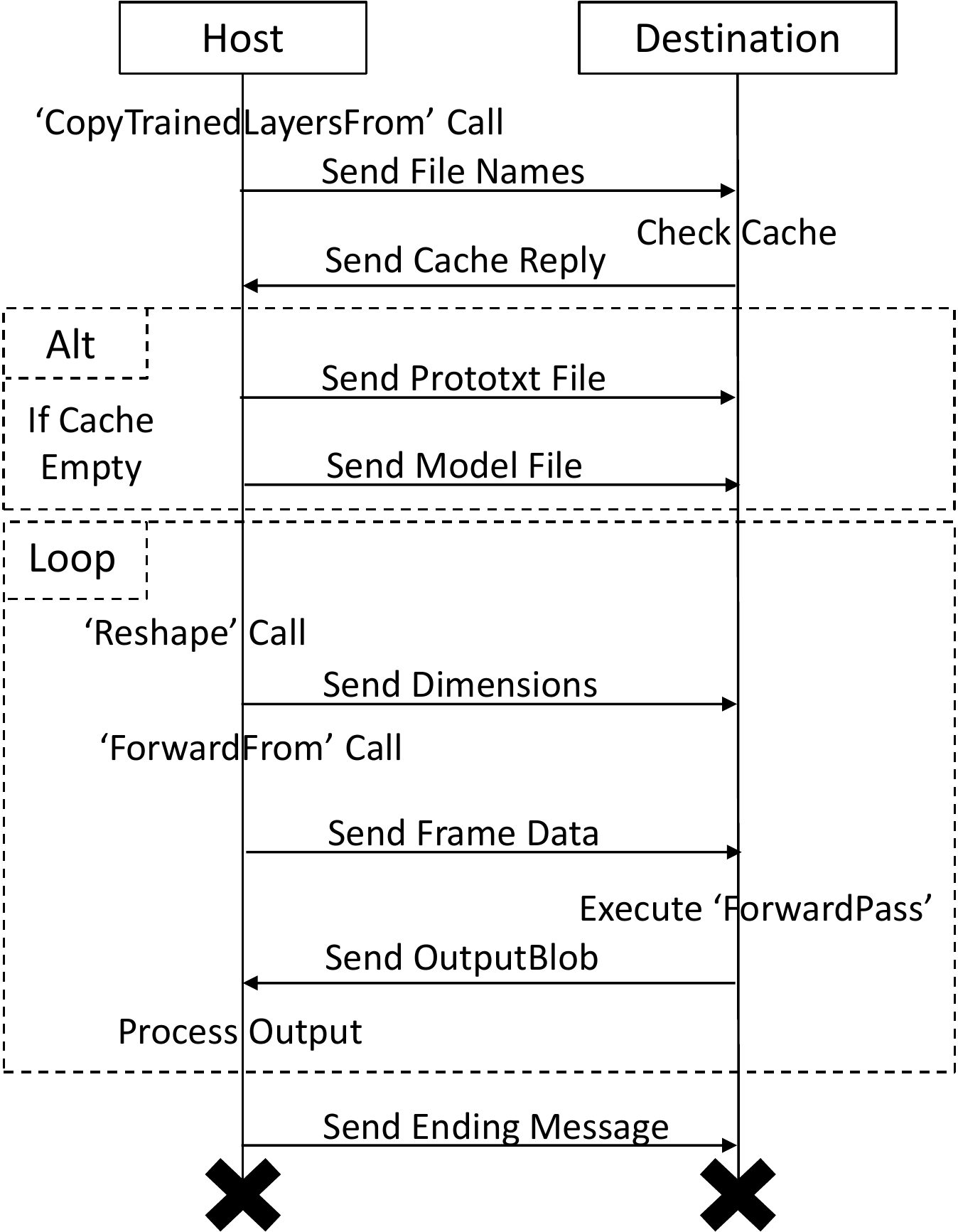}
  \caption{Sequence diagram of the \texttt{AVEC} framework}
  \vspace{-0.5cm}
  \label{fig:Sequence}
\end{figure}

This section presents the first prototype of the \texttt{AVEC} framework. 
\texttt{AVEC} aims to provide remote GPU acceleration for deep learning libraries, such as Caffe~\cite{caffe}. 
The Caffe library is open source and offers deep learning algorithms and models for C++ and Python. 
It uses the CUDA framework for acceleration on Nvidia GPUs. 
\texttt{AVEC} executes Caffe kernels within cloud-edge computing using an accelerator virtualization approach. 
\texttt{AVEC} is designed as a middleware that allows calls to the Caffe library to be executed in a remote accelerator by forwarding the input parameters of the GPU kernel to a remote destination in the edge or the cloud.

The \texttt{AVEC} framework is demonstrated on the Caffe deep learning library using the OpenPose application~\cite{OpenPose}. 
\texttt{AVEC} does not require that the source code of OpenPose is modified, rather operates as an interception library that runs outside the application level.

\begin{figure}[t]
  \centering
  \includegraphics[width=0.40\textwidth]{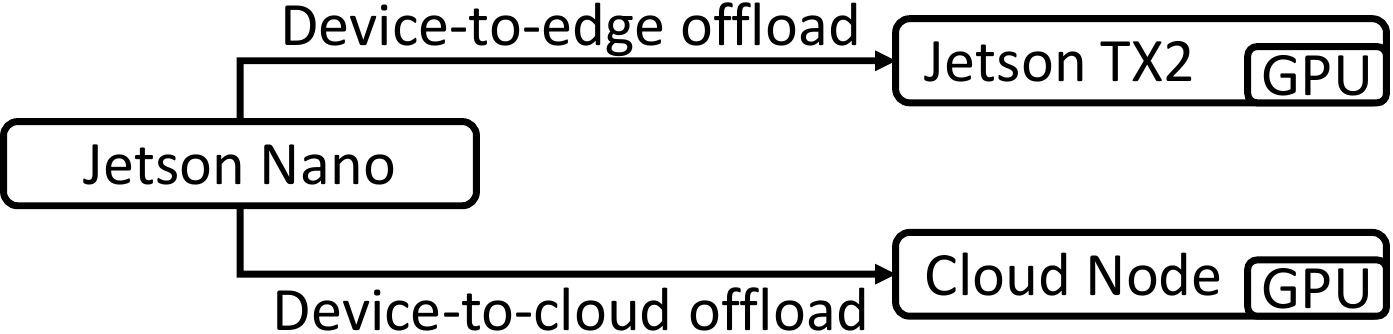}
  \caption{Organisation of experimental test-bed}
  \vspace{-0.25cm}
  \label{fig:Offloading}
\end{figure}

\begin{table}[t]
\centering
 \begin{tabular}{|l|c|c|c|} 
 \hline
  & Device & Edge Node & Cloud Node \\ [0.5ex] 
 \hline
 GPU Resource   & Jetson Nano & Jetson TX2 & RTX 2600\\
 \hline
 Flops (GPU) & 235 GFLOPS & 750 GFLOPS & 6.5 TFLOPS \\ 
 \hline
 \multirow{2}{*}{GPU Memory} & \multirow{4}{*}{4 GB LPDDR4} & \multirow{4}{*}{8 GB LPDDR4} & 6 GB \\
                             & \multirow{4}{*}{(Shared)} & \multirow{4}{*}{(Shared)} & GDDR6 \\
 \cline{1-1}\cline{4-4}
 \multirow{2}{*}{CPU Memory} & & & 32 GB \\
                             & & & DDR4 \\
 \hline
 GPU Cores & 128 & 256 & 1920 \\
 \hline
 CPU Cores & 4 & 4 & 8 \\
\hline
\end{tabular}
\caption{Hardware used in the lab-based test environment}
\vspace{-0.5cm}
\label{table:specComparison}
\end{table}

OpenPose takes media input, such as videos or images, and detects people within the frames and then highlights the pose of the detected person. Figure~\ref{fig:opDemo} shows an example output. Person and pose detection requires the execution of a GPU kernel using the CUDA language. \texttt{AVEC} aims to deploy the OpenPose accelerator kernel on a remote destination node if a GPU is not available or is resource constrained on a given device.

The architecture of the \texttt{AVEC} framework is shown in Figure~\ref{fig:architecture}. Its goal is to execute a GPU kernel from a host node in a remote accelerator in a destination node. 
For this \texttt{AVEC} extracts the data required to initiate the kernel in the destination node, runs it there and then sends the output of the kernel back to the host with the values needed to continue application execution. 
The framework employs a communication module to send data between the host and destination nodes. 
The original application on the host continues executing the application using the output of the remote kernel as if it was locally executed.

Figure~\ref{fig:Sequence} shows the sequence of activities between the host and destination when using OpenPose on the \texttt{AVEC} framework.
The communication between the host and destination is via TCP/IP using Boost ASIO~\cite{asio}. The host node in the cloud-edge environment may be a device or an edge node that does not have an accelerator or a sufficiently powerful accelerator. The host node executes the original application that requires the deep learning library (Caffe) and \texttt{AVEC}'s pre-loaded interception library to allow access to the virtualized GPU accelerator. When the application calls functions within Caffe, those calls are redirected to the interception library via API interception. The interception library creates a connection between the host and destination nodes (both in a cloud-edge network) using a TCP/IP connection. The destination node hosts a physical GPU and the original Caffe library. Once the destination node receives the forwarded requests, it executes the kernels required by the host node on the physical GPU of the destination node. The output of these functions is sent back to the user node in the same manner as they were received. When the output is received, the application continues executing on the host node. 

Within the Caffe library that OpenPose uses, the call to the Caffe kernels has been replaced with a function call to \texttt{AVEC}'s communications module. \texttt{AVEC} extracts arguments from the host and sends them to the destination. To this end, the Caffe \texttt{prototxt} file that contains the structure of the neural network must be initially sent to the destination. A caching mechanism is implemented so that these do not have to be sent for different executions of the kernel. In addition to the Caffe files, the input video frames to be analysed must also be sent with metadata regarding the frame, such as its size and aspect ratio. The process of assembling the frame into an array is carried out on the host to prevent unnecessary work being done in the destination node.

\section{Experimental Studies}
\label{sec:Experiments}
\begin{table}
\centering
 \begin{tabular}{|c|c|c|c|} 
 \hline
 Number of Images & Cloud Node & Edge Node & Device \\ [0.5ex] 
 \hline
 64 & 8.13 & 69.47 & 130.77 \\ 
 \hline
 128 & 13.82 & 134.02 & 256.64 \\
 \hline
 256 & 25.98 & 258.19 & 497.06 \\
 \hline
 \end{tabular}
  \caption{Execution time (s) of the OpenPose application}
  \vspace{-0.25cm}
  \label{table:openPoseBase}
\end{table}%

\begin{table}
\centering
 \begin{tabular}{|c|c|c|c|} 
 \hline
  & Device & Edge Node & Cloud Node \\ [0.5ex] 
 \hline
  Model transfer time (s) & 6.43 & 5.937 & 1.757 \\ 
 \hline
 \end{tabular}
  \caption{Time taken to transfer COCO model to the GPU}
  \vspace{-0.5cm}
  \label{table:GPUCOPY}
\end{table}%

This section presents experimental studies on \texttt{\texttt{AVEC}} to implement GPU virtualization in a cloud-edge environment. The capabilities of \texttt{AVEC} are tested using the OpenPose application. 

\subsubsection{Experimental Setup}
A lab-based test environment comprising resources that are comparable to a user device, an edge node and a cloud resource are setup as shown in Table~\ref{table:specComparison}.
Experiments are carried out with a one to one connection, such that the host device (Jetson Nano) will only offload to one destination node during the execution, as shown in Figure~\ref{fig:Offloading}.

Tests are carried out on images and videos. 
The COCO dataset~\cite{COCOmicrosoft} is used to test images (2017 version). For videos, a sample provided by OpenPose is used. The images chosen from the COCO dataset are randomly selected and set into batches of 64, 128, and 256 images. These images vary in number of objects (i.e. persons) present in them (images that do not have persons are also present). The same batches of images are used for all tests to ensure that the results are comparable. OpenPose will superimpose the pose of persons present on the image as shown in Figure~\ref{fig:opDemo}.

\begin{figure*}
     \centering

     \begin{subfigure}[b]{0.48\textwidth}
       \centering
       \includegraphics[width=\textwidth]{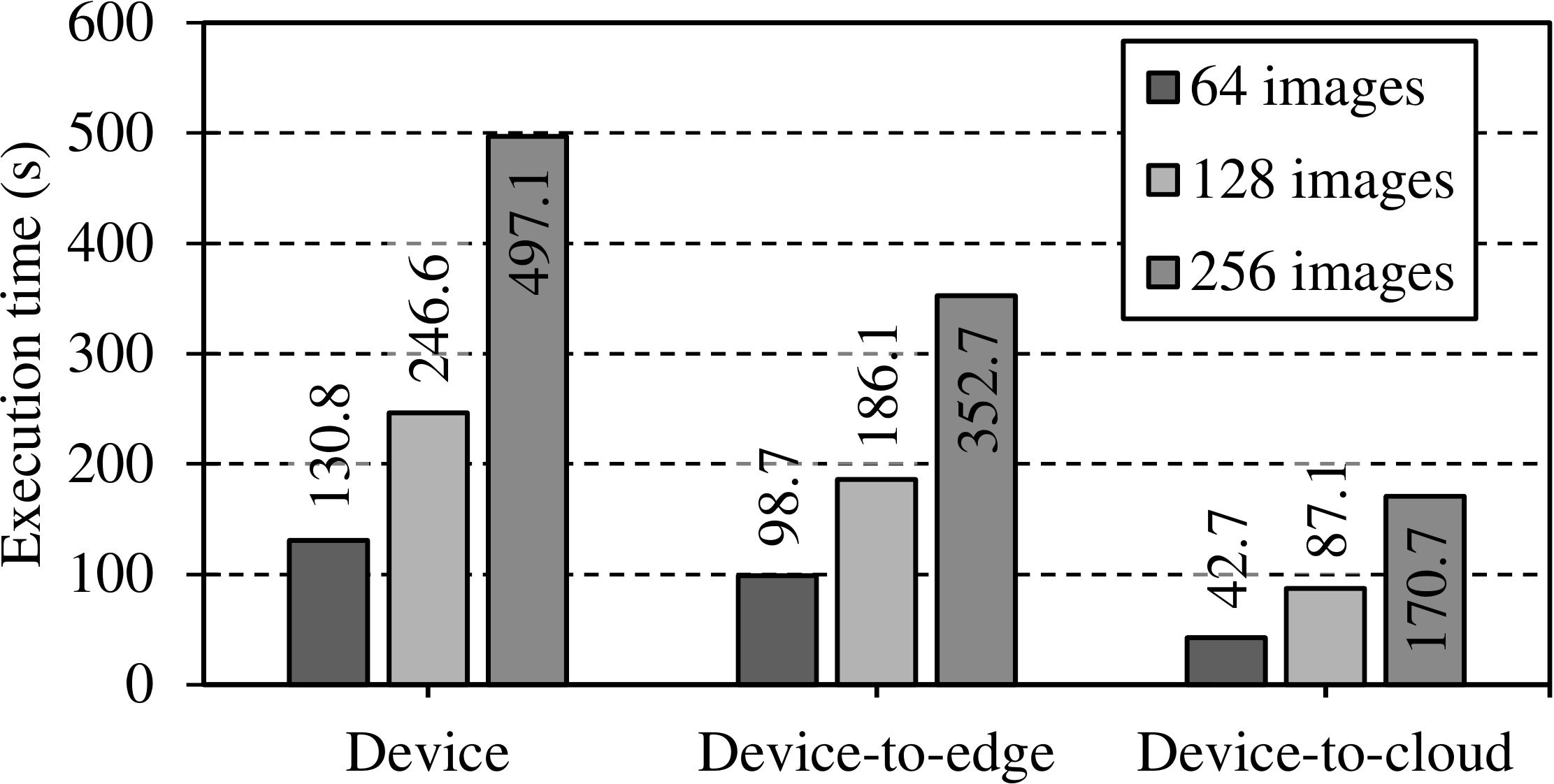}
       \caption{Image batches}
       \label{fig:tests_images}
     \end{subfigure}
\hfill
     \begin{subfigure}[b]{0.48\textwidth}
       \centering
       \includegraphics[width=\textwidth]{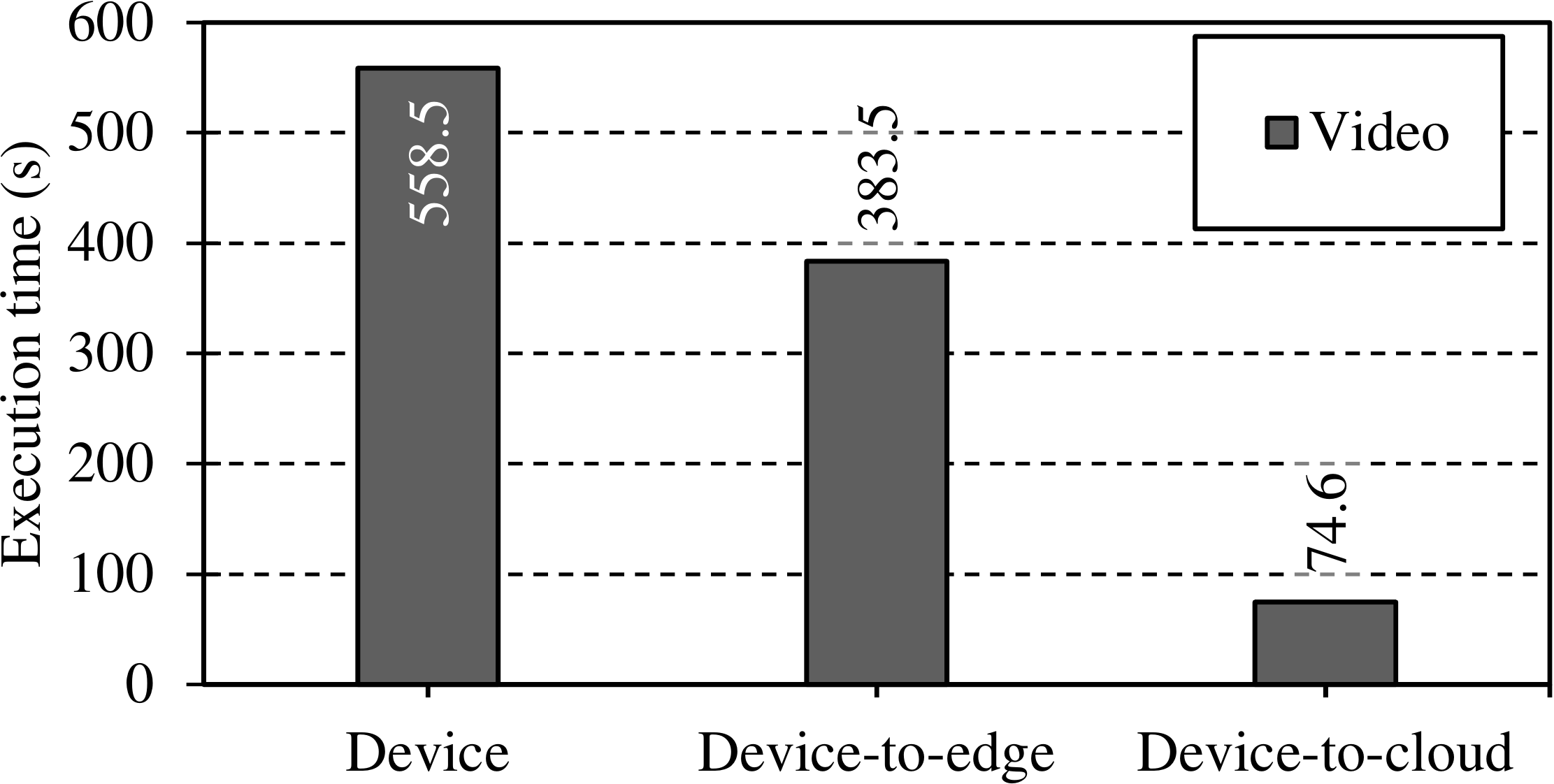}
       \caption{Video}
       \label{fig:Video}
     \end{subfigure}

     \caption{Execution time for different image batches and video when using \texttt{AVEC}}
     \vspace{-0.5cm}
     \label{fig:ExecutionTimes}
\end{figure*}

\subsubsection{\texttt{AVEC} for Images}
The first tests were carried out using the randomly selected batches of 64, 128 and 256 images with different resolutions. Each time a selection of images was taken from the COCO dataset, a different seeding was used so that when the random selection of images to sort into each batch took place, the same images would not be placed in all image batches. This was done to ensure that a variety of images were used across the different tests. The original OpenPose application was timed and tested with the image batches and the results are shown in Table~\ref{table:openPoseBase}. It is observed that the time taken to process each image batch is approximately twice the time taken for a previous batch. 

The same images were executed using \texttt{AVEC} on the same image batches as in the above test. The results shown in Figure~\ref{fig:tests_images} indicate that an increase in speed is achieved as \texttt{AVEC} offloads the computationally intensive components of OpenPose to remote GPUs. The time is reduced by a significant amount when the edge node is used, but when the device offloads to the cloud node a substantial speedup is observed. This is as expected, since the resources available on the GPU increases from the device through to the edge and the cloud. 

Table~\ref{table:GPUCOPY} shows the time taken to copy the deep learning model to the GPU. The model needs to be transferred to the GPU. In the experiments, the COCO model requires up about 5.5GB of memory on the GPU and is copied once during the initialisation to the GPU and remains there throughout execution.

\subsubsection{\texttt{AVEC} for Videos}
The video used for this test is an 8 seconds clip consisting of 204 frames with a resolution of 368~x~656 pixels. It has a large number of objects (i.e. persons) present in each frame, therefore, rendering is required on the GPU. The results obtained from offloading are shown in Figure~\ref{fig:Video}. A similar trend as seen for the images is noted, and \texttt{AVEC} provides larger amount of improvement with the cloud node than with the edge node. The speedup obtained on both images and video using \texttt{AVEC} is shown in Table~\ref{table:SPEEDUP}. It is observed that \texttt{AVEC} can deliver up to 7.48x speedup in the test with the video when offloading the workload to the cloud.

\begin{table}[t]
\centering
 \begin{tabular}{|c|c|c|} 
 \hline
 Test type & Device-to-edge offload & Device-to-cloud offload \\ [0.5ex] 
 \hline
  64 Images & 1.32x & 3.06x \\ 
 \hline
  128 Images & 1.32x & 2.83x \\ 
 \hline
  256 Images & 1.40x & 2.91x \\ 
 \hline
 Video & 1.45x & 7.48x \\  
 \hline
 \end{tabular}
  \caption{Speedup for images and video using \texttt{AVEC}}
  \vspace{-0.5cm}
  \label{table:SPEEDUP}
\end{table}%

Another dimension to quantify the performance benefit of \texttt{AVEC} is by measuring the frames per second~(FPS) processed. This shown in Table~\ref{table:FPS}. As expected, an improvement in FPS is noted when offloading to more powerful GPUs. However, the edge is still quite limited in its computational capabilities and therefore there is limited acceleration.

Another observation is that the video performs better than the images. This may be due to the video not being saved to memory after each frame is processed, whereas each processed image is saved with the OpenPose prediction transposed onto it.

\begin{table}[t]
\centering
 \begin{tabular}{|c|c|c|c|c|c|} 
 \hline
 Test & Device & \multirow{2}{*}{Edge} & \multirow{2}{*}{Cloud} & Device-to-edge & Device-to-cloud\\ 
 type &        &      &       & using \texttt{AVEC} & using \texttt{AVEC}\\ 
 \hline
 Images & 0.5 & 1.1 & 10.5 & 0.65 & 2 \\ 
 \hline
  Video & 0.4 & 0.7 & 9 & 0.6 & 3.1 \\
 \hline
 \end{tabular}
  \caption{Comparing frames per second processed on the device, edge, and cloud resource natively and when offloading from the device to the edge and to the cloud using \texttt{AVEC}}
  %\vspace{-0.5cm}
  \label{table:FPS}
\end{table}%

\subsubsection{Profiling \texttt{AVEC} performance}
The Nvidia profiler (nvprof) was used to measure the time spent for computation of the GPU kernels, data transfers (i.e. communication), and other overheads. It is noted that \texttt{AVEC} reduces the execution time on the GPU when offloading, but incurs overheads due to transferring GPU kernel arguments and results between nodes. 

When running the OpenPose application natively without the use of \texttt{AVEC}, 27 GPU kernels are used through the execution sequence. Using \texttt{AVEC}, 13 kernels are executed on the host device and 17 kernel executions on the destination node (edge or cloud). The three additional kernels are CUDA calls required on both the host and destination. The host retains all OpenPose related kernels (e.g. renderPoseCoco), but the destination node executes all Caffe related kernels. 

\begin{figure}[t]
  \centering
  %\vspace{-0.5cm}
  \includegraphics[width=0.48\textwidth]{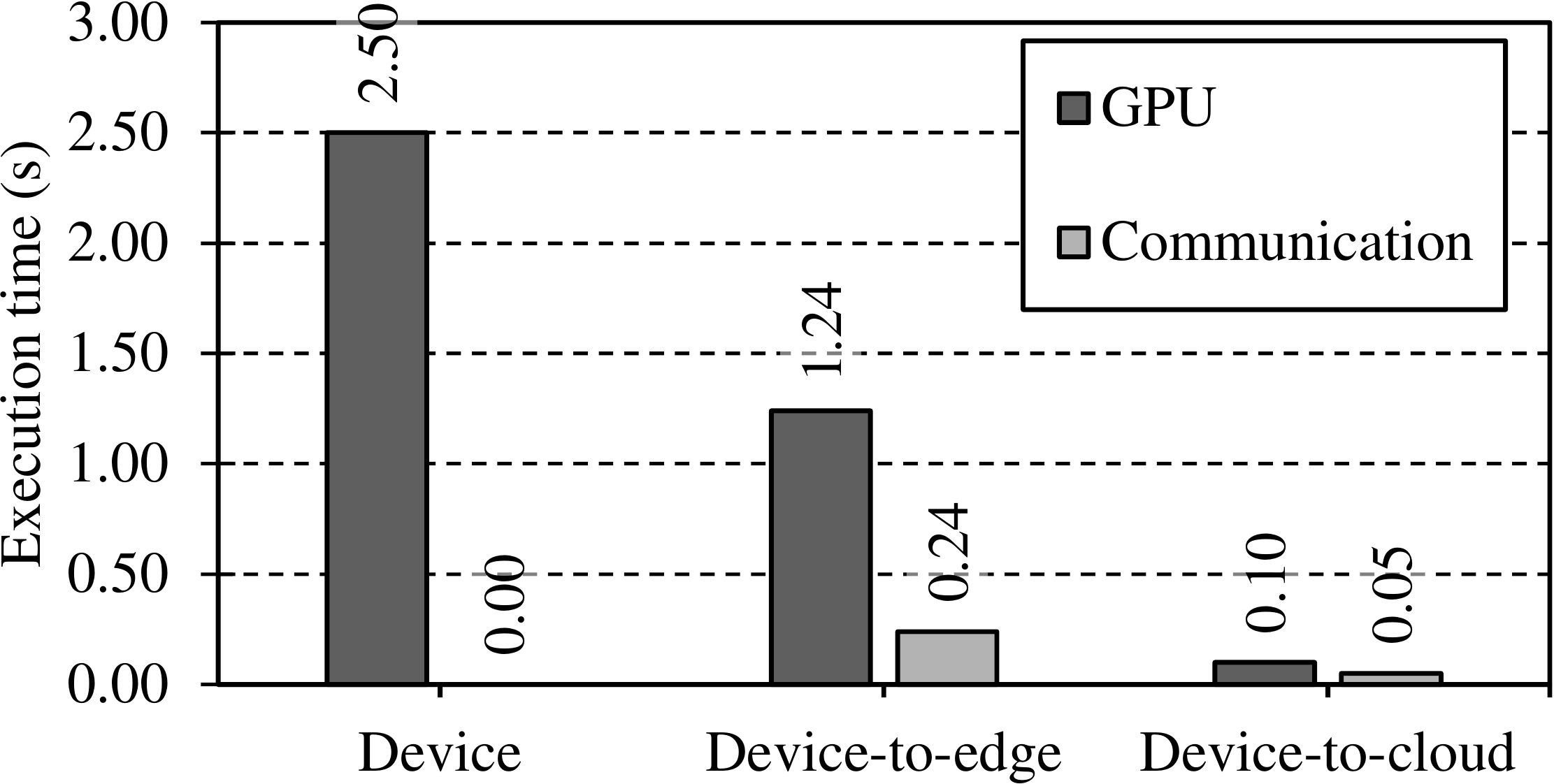}
  \caption{Profiling single frame execution when using \texttt{AVEC}}
  \label{fig:breakdown}
\end{figure}

\begin{figure*}
     \centering

     \begin{subfigure}[b]{0.48\textwidth}
       \centering
       \includegraphics[width=\textwidth]{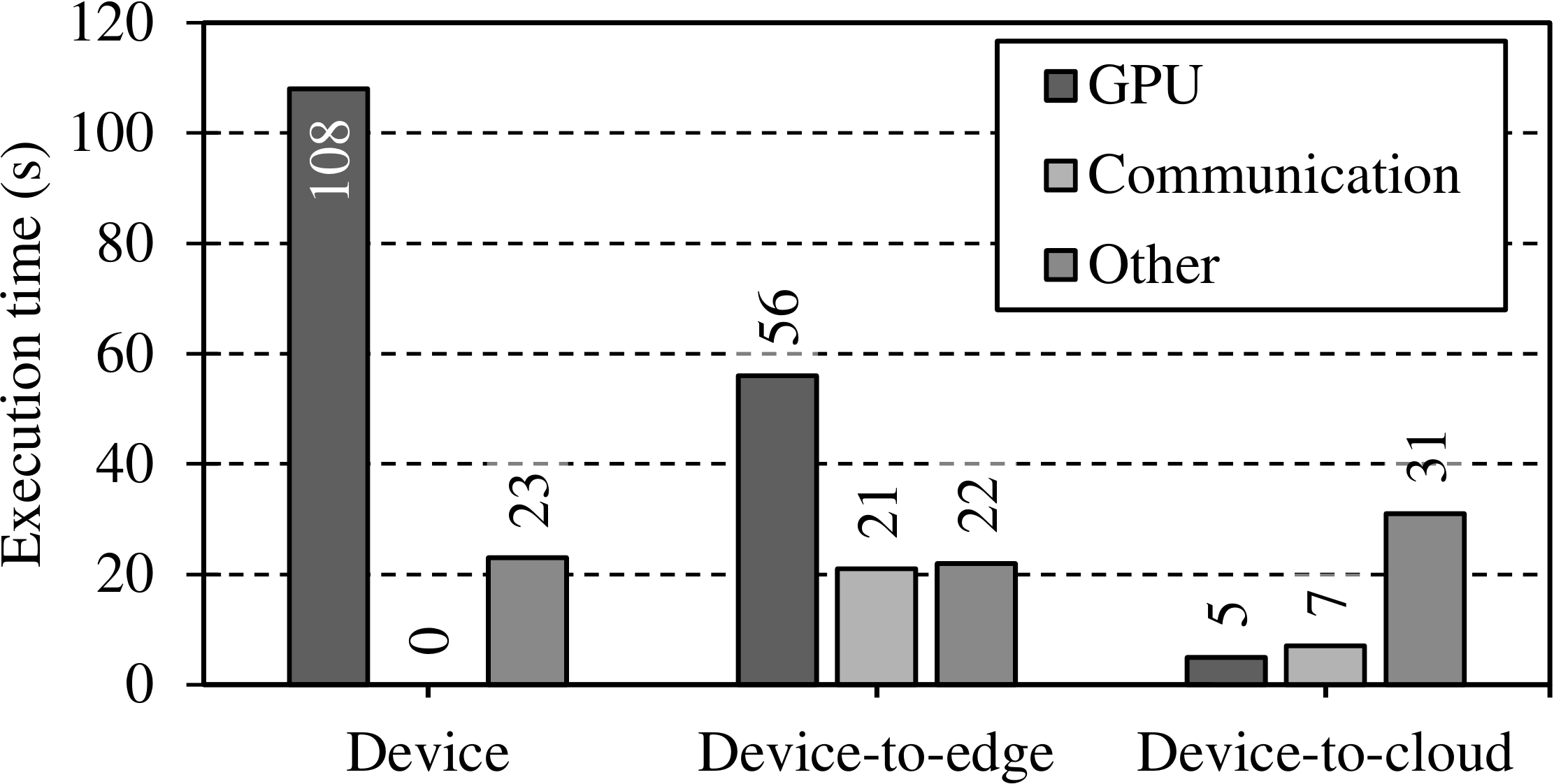}
       \caption{64 images}
       \label{fig:64Overheads}
     \end{subfigure}
\hfill
     \begin{subfigure}[b]{0.48\textwidth}
       \centering
       \includegraphics[width=\textwidth]{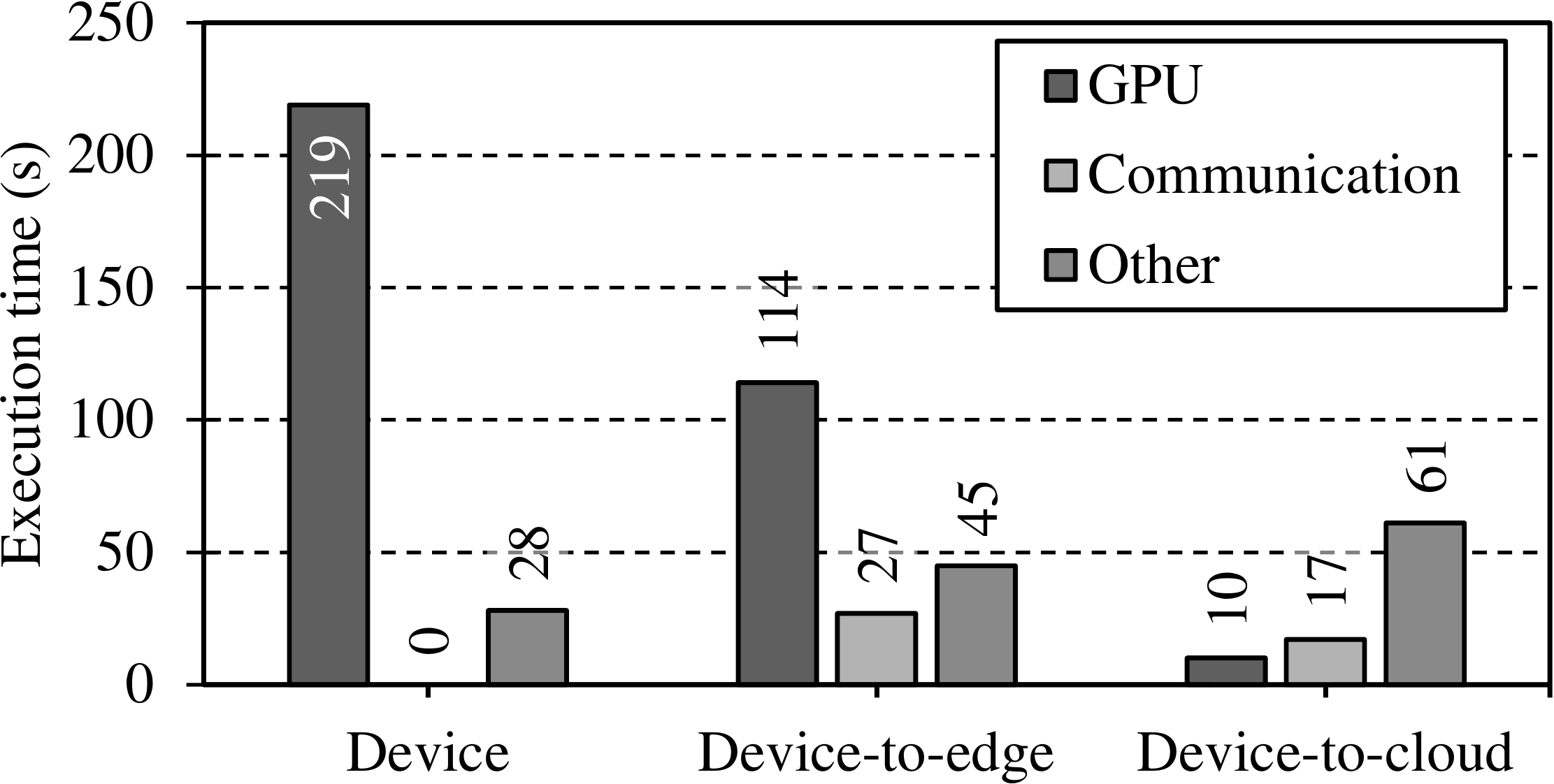}
       \caption{128 images}
       \label{fig:128Overheads}
     \end{subfigure}
\hfill
     \begin{subfigure}[b]{0.48\textwidth}
       \centering
       \includegraphics[width=\textwidth]{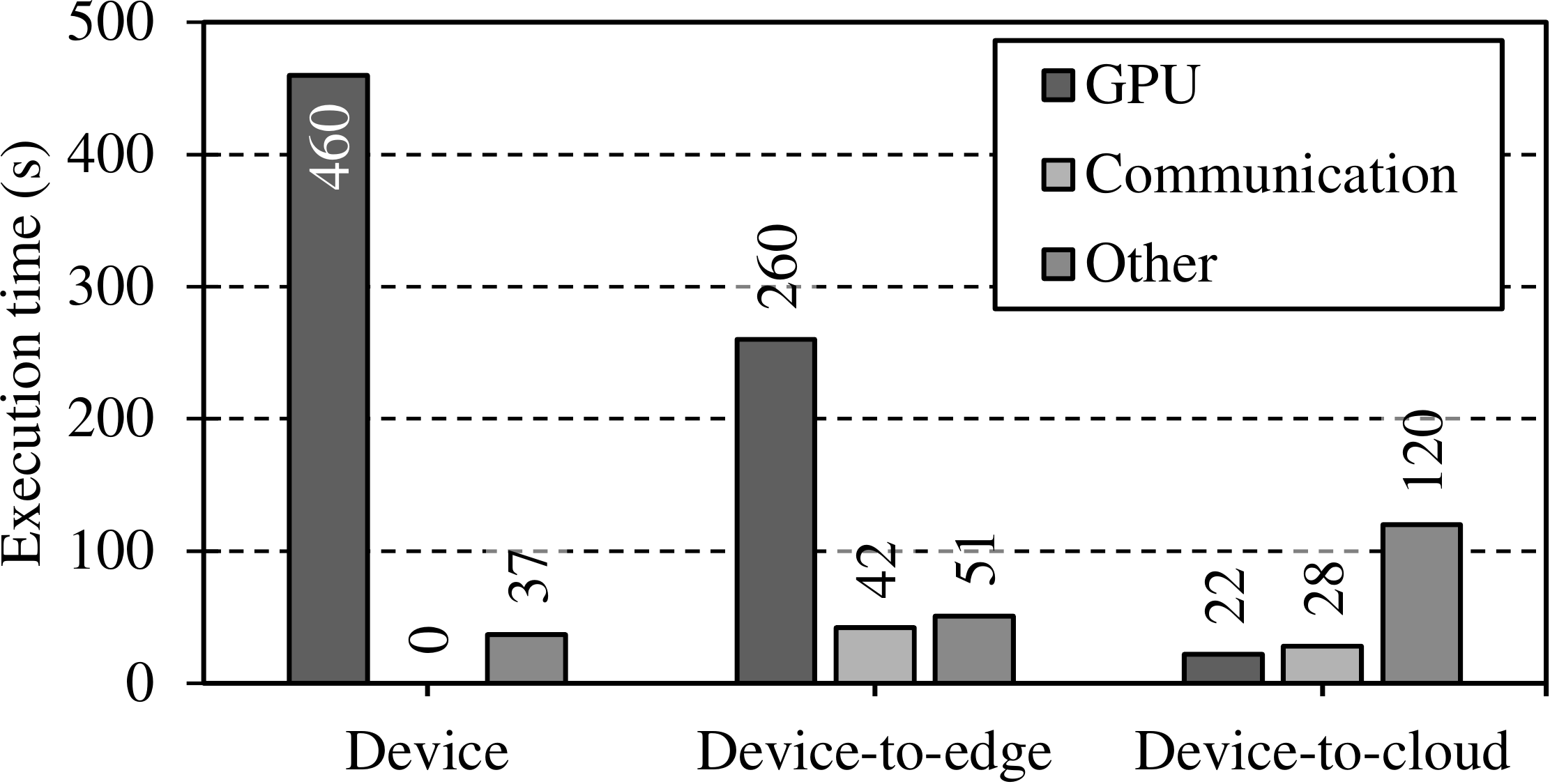}
       \caption{256 images}
       \label{fig:256Overheads}
     \end{subfigure}
\hfill
     \begin{subfigure}[b]{0.48\textwidth}
       \centering
       \includegraphics[width=\textwidth]{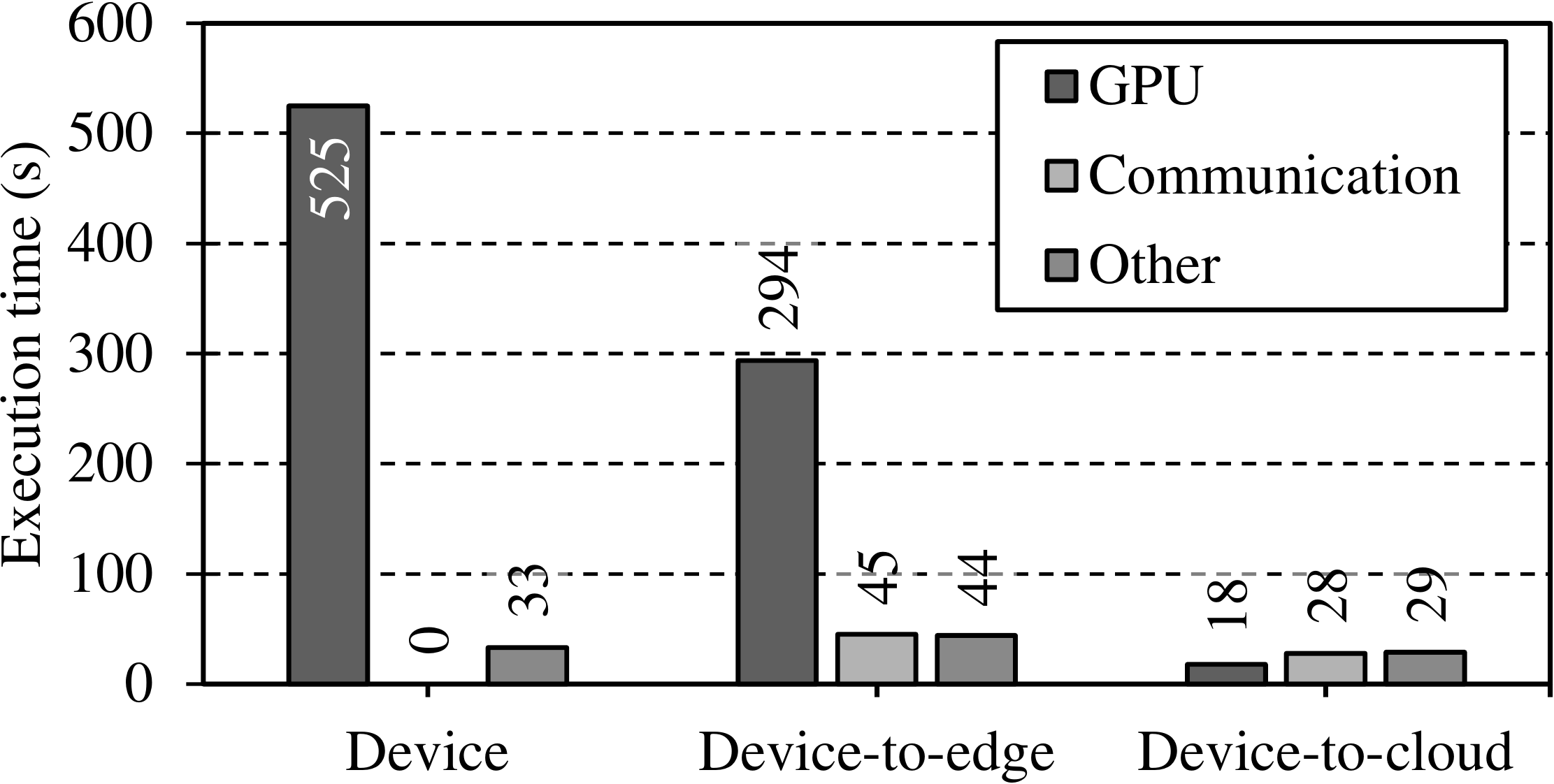}
       \caption{Video}
       \label{fig:VideoOverheads}
     \end{subfigure}

     \caption{Profiling of different image batches and video when using \texttt{AVEC}. `GPU' refers to time taken for execution on the GPU, `Communication' refers to time taken to send data between nodes, and `Other' refers to time spent on OpenPose specific tasks carried by the CPU on the host device.}
     \vspace{-0.5cm}
     \label{fig:Overheads}
\end{figure*}

The time taken to complete a forward pass, and thus make a real time pose estimation depends on the frame resolution (larger frames take a longer time to process). The results discussed below are from using the sample video used in previous experiments.

One execution cycle is defined as the time taken to (i)~send the frame data as float, (ii)~send the resolution as an array, (iii)~send the frame size as an integer, (iv)~execute the forward pass and return the results. The frame width variable is two integers (8 bytes). The frame size is a single integer (4 bytes). The frame size will be the dimensions of the frame (1~x~3~x~368~x~656). Finally, the output result sent back to the host node will always be the dimensions of the frame divided by a constant \texttt{c}. This constant depends on the model used (in this case the constant for the model used is 3.368421). We can formalise the amount of data transferred (DT) per frame as shown in Equation~\ref{equation:X}. In these experiments, the amount of data transferred per frame is approximately 3.75~MB.
 %\[Data Transferred = (2 * 4) + (1 * 4) + (Dimensions * 4) + ((Dimensions / c) * 4)]
 \begin{equation}
  DT = (2 * 4) + (1 * 4) + (Dims * 4) + ((Dims / c) * 4)
  \label{equation:X}
 \end{equation}

Figure~\ref{fig:breakdown} provides a breakdown of the profiled data when natively executing on the device and when offloading using \texttt{AVEC}.
When considering device-to-cloud offloading, on average each frame requires 0.15 seconds of processing time for a full execution cycle. Approximately 0.1 seconds are spent in the GPU executing kernels to carry out the forward pass operation. An overhead of 0.05 seconds is incurred in transferring data between the device and the cloud. When considering device-to-edge offloading approximately 1.48 seconds are required per frame and the time taken in the GPU of the edge node is 1.24 seconds. An overhead of 0.24 seconds is incurred by transferring data. The time taken for the host device to complete its own forward pass is around 2.5 seconds.

A breakdown of the time spent by \texttt{AVEC} on different image batches and video is shown in Figure~\ref{fig:Overheads}. Across all test types it is noted that the GPU execution time is reduced with \texttt{AVEC}. When offloading to the cloud node the communication overhead is larger than the execution time on the GPU. As expected, the communication overhead incurred by the edge is greater than the cloud although the same amount of data is transferred. This is due to the differences in hardware, resulting in the Communication Module of \texttt{AVEC} being executed faster on the CPU of the cloud node than on the edge. 

\section{Conclusions}
\label{sec:Conclusion}
In this paper is presented a first prototype of the \texttt{AVEC} framework (accelerator virtualization in cloud-edge computing). The research investigated the potential of GPU accelerator virtualization to improve the performance of deep learning workloads in cloud-edge environments. More specifically three research questions were addressed: (Q1) How can accelerator virtualization be offered at the edge given its limited compute resources? (Q2) What overheads are incurred in virtualizing accelerators at the edge? (Q3) Can GPU accelerator virtualization at the edge improve the execution performance of real-world applications?

\texttt{AVEC} is underpinned by a virtualization technique that employs API interception and forwarding, that requires no source code modification of applications. A thorough evaluation of the overheads that are incurred in using \texttt{AVEC} was performed and its impact on the execution of workloads was analysed. A real world use-case was employed, namely the OpenPose application using the Caffe deep learning library. OpenPose is a computationally intensive application that low powered embedded devices cannot accelerate. However, performance acceleration was achieved when using \texttt{AVEC}. It was noted that although there are communication overheads in transferring data from devices to a remote GPU, \texttt{AVEC} still delivers a speedup reaching up to a maximum of 7.48x. The first implementation of \texttt{AVEC} has the following limitations providing opportunities for further research:

(i) \textit{Only considers deep learning libraries}: 
Currently, \texttt{AVEC} only incorporates the Caffe deep learning library. In the future, it is envisioned that \texttt{AVEC} will incorporate other deep learning libraries to deliver a more comprehensive remote accelerator virtualization solution and thus enable \texttt{AVEC} to become more mature.

(ii) \textit{No mechanism to enable migration of workloads between accelerators}: 
Due to the dynamic nature of cloud-edge environments and its benefit for mobile users, \texttt{AVEC} must be capable of moving workloads to alternate locations based on the requirements of users or their workloads. This is implemented by packaging workloads into deployable containers and moving them across accelerators in a cloud-edge network. Migration of workloads will also enhance fault-tolerance in \texttt{AVEC} in the event of nodes that host accelerators failing.

(iii) \textit{Lack of device-aware scheduling}:
Given the heterogeneous nature of cloud-edge computing, \texttt{AVEC} needs to be aware of the accelerators that are available within the network that are capable of running a given workload. Moreover, metrics such as accelerator usage, power usage or latency of these accelerators should be taken into account for scheduling a given workload.

\section*{Acknowledgements}
We gratefully acknowledge the support of NVIDIA Corporation with the donation of the Jetson TX2 Development Kit used for this research.

%\newpage
\bibliographystyle{IEEEtran}
\bibliography{IEEEabrv, references}

\clearpage
\end{document}